\documentstyle[prl,aps,psfig]{revtex}
\begin{document}
\draft
\twocolumn[\hsize\textwidth\columnwidth\hsize\csname @twocolumnfalse\endcsname
\title{Anomalous modes drive vortex dynamics in confined Bose-Einstein
condensates}
\author{David L. Feder,$^{1,2}$ Anatoly A. Svidzinsky,$^3$
Alexander L. Fetter,$^3$ and Charles W. Clark$^2$}
\address{$^1$Clarendon Laboratory, University of Oxford, Parks Road, Oxford
OX1 3PU, U.K.}
\address{$^2$Electron and Optical Physics Division, National Institute of
Standards and Technology, Technology Administration, U.S. Department of
Commerce, Gaithersburg, MD, 20899-8410}
\address{$^3$Laboratory for Advanced Materials and Department of Physics,
Stanford University, Stanford, CA 94305-4045}
\date{\today}
\maketitle
\begin{abstract}
The dynamics of vortices in trapped Bose-Einstein condensates are investigated
both analytically and numerically. In axially symmetric traps, the
critical rotation frequency for the metastability of an isolated vortex
coincides with the largest vortex precession frequency (or anomalous mode)
in the Bogoliubov excitation spectrum. As the condensate
becomes more elongated, the number of anomalous modes increases.  The largest
frequency of these modes exceeds both the thermodynamic critical frequency
and the nucleation frequency at which vortices are created dynamically. Thus,
anomalous modes describe not only the critical rotation frequency for creation
of the first vortex in an elongated condensate but also the vortex precession
in a single-component spherical condensate.

\end{abstract}
]

\narrowtext

Bose-Einstein condensation (BEC) and superfluidity are two entangled core
issues of low-temperature physics, and recent experimental developments now
allow us to study them in a nearly ideal system: a dilute gas of alkali atoms
with well-understood interactions~\cite{reviews,Varenna,others}. Many key
aspects of BEC have been clarified since the first experimental observations,
and during the past year much attention has been given to manifestations of
superfluidity. The `scissors mode' of excitation of trapped
condensates~\cite{scissors} implies the irrotational flow characteristic of
superfluids. Quantized vortices, which have long been known as fundamental
excitations of superfluid helium and superconductors, have been observed
directly in one-~\cite{Dalibard,Anderson} and two-component~\cite{Matthews}
trapped Bose gases.

Multiple vortices have recently been generated in
confined single-component condensates~\cite{Dalibard} by rotating a weakly
anisotropic trap at an angular frequency $\Omega$. This approach resembles
the classic rotating-bucket experiments on liquid helium~\cite{Donnelly},
where vortices first appear~\cite{Packard} at a critical
frequency $\Omega_c$ above which the presence of the vortex lowers the total
free energy~\cite{Packard1,Fetter1}.  There is a discrepancy, however,
between the observations of Madison {\it et al.}~\cite{Dalibard} and
simple theoretical considerations based on the equilibrium
solution of the Gross-Pitaevskii (GP) equation, which has otherwise
been remarkably successful in describing condensate behavior~\cite{reviews};
in particular, the frequency at which vortices are first observed is
significantly higher than equilibrium estimates of $\Omega_c$.

The resolution of this discrepancy is the subject of the present paper.
In particular, we find that the `anomalous modes' of the Bogoliubov spectrum
determine the critical rotation frequencies at which vortex
arrays are observed in the ENS experiment~\cite{Dalibard}; 
they also determine the frequency of precession
of the vortex core observed at JILA~\cite{Anderson}.

In trapped condensates, several factors influence
the critical rotation frequency for appearance and stability
of vortices, and for their subsequent dynamical motion:

(1) An energy barrier at the surface of the condensate~\cite{Fetter2}
inhibits vortex penetration of its interior. Thus,
vortices are predicted to nucleate spontaneously at a frequency
$\Omega_{\nu}>\Omega_c$~\cite{Feder1}, whose value coincides
with the criterion for the Landau instability of surface
excitations~\cite{Dalfovo,Pethick,Isoshima} and 
with the effective disappearance of
the surface barrier~\cite{Feder1}. We show below that, 
while $\Omega_{\nu}$
exceeds $\Omega_c$ for the cigar geometry of the
ENS experiments~\onlinecite{Dalibard}, it is
still lower than the frequency at which vortices are 
first observed.

(2) Stringari~\cite{Stringari} has shown that $\Omega_c$ increases with $T$,
reaching a maximum close to the BEC transition $T_c$. The temperature $T\sim
0.8T_c$ required to match the observed critical frequency, however, is much
larger than the temperature at which the experiments are performed.

(3)   A condensate with a vortex first becomes stable against
small perturbations at the metastability frequency $\Omega_m$. One
definition for
metastability is that the energy per particle must not decrease
under infinitesimal displacements of the vortex from the condensate center;
for a disk-shaped condensate, this yields $\Omega_m={\case3/5}\Omega_c$ in the
Thomas-Fermi (TF) limit~\cite{Fetter2}. 
Equivalently, the Bogoliubov excitation spectrum must
contain only modes with positive energy. A non-rotating
axisymmetric condensate with a singly quantized vortex has
at least one ``anomalous mode''; such modes have a negative 
excitation energy $\varepsilon_a$ and a
positive norm~\cite{Dodd,Rokhsar}. For a rotating axisymmetric
trap, these anomalous  modes are Doppler-shifted upward 
by $\Omega$, with the
metastability frequency $\Omega_m=\mbox{max}(|\varepsilon_a|)/\hbar$. As
shown below, $\Omega_m$ can exceed both $\Omega_c$ and $\Omega_{\nu}$ in cigar
traps.

This criterion for the onset of linear stability agrees well with the large
frequency at which vortices first appear experimentally in cigar-shaped
condensates~\cite{Dalibard}. It
also explains why the observed critical frequency is independent of whether
the condensate is first cooled and subsequently rotated, or {\it vice versa}.
Otherwise, the critical rotation should be greater for the cool-and-then-rotate
scenario, since vortices must surmount the energy barrier at the condensate
surface, penetrating the cloud only at the nucleation frequency
$\Omega_{\nu}>\Omega_c$.

(4) The anomalous modes correspond to the precession of the vortex
core about the origin of the trap~\cite{Fetter2,Linn}. For $\Omega<\Omega_m$ ,
the vortex moves in the same direction as the superfluid flow around the
core. Hence,  the anomalous mode(s) should describe the motion of one-component
vortices in recent JILA experiments~\cite{Anderson,Matthews}.

The present work links anomalous modes with these various observations. We
proceed from the zero-temperature description of the order parameter (or
wave function) for a condensate of $^{87}$Rb atoms, as a solution
time-dependent GP equation

\begin{equation}
i\partial_{t}\psi({\bf r},t)
=\left(T+V_{\rm trap}+V_{\rm H}-\Omega L_z\right)\psi({\bf r},t),
\label{gp}
\end{equation}

\noindent with the kinetic energy operator given by
$T=-{\case1/2}\vec{\nabla}^2$, the Hartree field by
$V_{\rm H}=4\pi\eta|\psi|^2$, and trap potential by
$V_{\rm trap}={\case1/2}\left[\lambda^2(1+\epsilon_x)x^2
+\lambda^2(1+\epsilon_y)y^2+z^2\right]$, where 
$\lambda=\omega_{\rho}/\omega_z$ is
the trap anisotropy and $\epsilon_x$, $\epsilon_y$
describe (small) departures of the trap from
axial symmetry. The centrifugal term $-\Omega
L_z=i \Omega \left(x\partial_y-y\partial_x\right)$ 
appears for systems
rotating about the $z$ axis at a frequency $\Omega$. 
Energy, length, and
time are respectively
given in harmonic oscillator units $\hbar\omega_z$,
$d_z=\sqrt{\hbar/M\omega_z}$, and $\omega_z^{-1}$, where
$\omega_z$ is the axial trap frequency and $M$ is the atomic mass.
Normalizing the condensate wave function $\int d{\bf r}|\psi({\bf r},t)|^2=1$
yields the scaling parameter $\eta=Na/d_z$, 
where $a=100a_0\approx 5.29$~nm is the
$^{87}$Rb scattering length~\cite{Eite} and $N$ is the number of
condensate atoms. As parameters typical of the recent experiments at
ENS~\cite{Dalibard} and JILA~\cite{Anderson,Matthews}, we take
$N=1.4\times 10^5$ with $(\nu_{\rho},\nu_z)=(\omega_{\rho},\omega_z)/2\pi
=(169,11.7)$~Hz and $(\epsilon_x,\epsilon_y)=(0.03,0.09)$, and $N=3\times 10^5$
with $(7.8,7.8)$~Hz and $\epsilon_x=\epsilon_y=0$, respectively.

The stationary solutions of the GP equation, defined as local minima of the
free energy, are found numerically by norm-preserving imaginary time
propagation using an adaptive stepsize Runge-Kutta integrator. The spatial
dependence of the complex condensate wave function employs a discrete-variable
representation~\cite{Feder2} based on Gaussian quadrature, and is assumed
to be even under inversion of $z$. The numerical techniques are described in
greater detail elsewhere~\cite{Feder1,Feder2}. The initial condensate amplitude
is taken to be the TF wave function
$\psi_{\rm TF}=\sqrt{(\mu_{\rm TF}-V_{\rm trap})/4\pi\eta}$, where $\mu_{\rm
TF}={\case1/2}(15\lambda^2\eta)^{2/5}$ is the chemical potential.
The GP equation for a given value of $\Omega$ and $N$ is
propagated in imaginary time until the fluctuations in both the chemical
potential and the norm become smaller than $10^{-11}$.

Once the system reaches equilibrium, its response to
small disturbances is found by substituting
$\psi\to e^{-i\mu t}\left(\psi+ue^{-i\varepsilon t}+ve^{i\varepsilon t}\right)$
in Eq.~(\ref{gp}). The Bogoliubov spectrum of
eigenvalues $\varepsilon$ is found by linearizing in the 
quasiparticle amplitudes
$u({\bf r})$ and $v({\bf r})$.

\begin{figure}[tb]
\psfig{figure=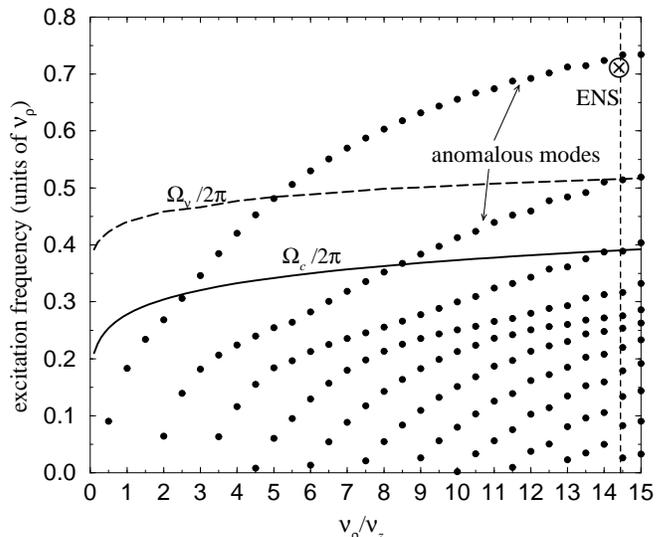,width=\columnwidth,angle=270}
\caption{The numerical values for the thermodynamic critical frequency
$\Omega_c/2\pi$ (solid curve), the nucleation frequency $\Omega_{\nu}/2\pi$
(dashed curve), and the even $z$-parity anomalous mode frequencies
(circles) are
shown as a function of trap anisotropy $\lambda=\nu_{\rho}/\nu_z$.
Parameters are
$N=1.4\times 10^5$,
$\nu_z=11.7$~Hz, and $\epsilon_x=\epsilon_y=0$. The dotted vertical line
corresponds to anisotropy relevant to the ENS experiments, and the
`$\otimes$' labels the frequency at which vortices are first observed.}
\label{omega_ani}
\end{figure}

Figure~\ref{omega_ani} shows the critical frequencies
$\Omega_c/2\pi$ and $\Omega_{\nu}/2\pi$ as a function of trap anisotropy
$\lambda$. We ignore the small in-plane trap
distortion (setting $\epsilon_x=\epsilon_y=0$) 
in order to simplify the computation.
In this axisymmetric system, the thermodynamic critical frequency
$\Omega_c$ is the energy difference between condensates with and without a
vortex at the trap center (divided by $\hbar$), since the rotation Doppler
shifts the former by exactly $-\Omega$; for the parameters
of the ENS experiment, $\lambda\approx 14.44$,
we obtain
$\Omega_c/2\pi\approx 0.4\nu_{\rho}\approx 68$~Hz, 
in good agreement with the
TF estimate but much lower than the observed value
$\Omega_{\rm obs}/2\pi\approx 0.7\nu_{\rho}\approx 120$~Hz. We find
that inclusion of the small in-plane anisotropy does not 
appreciably increase
the value of $\Omega_c$, in contrast with results reported
recently~\onlinecite{Garcia}. The critical nucleation frequency
$\Omega_{\nu}=\mbox{min}(\varepsilon_{nm}/\hbar m)$ defines the rotation
frequency at which the first Bogoliubov excitation of the vortex-free
condensate becomes unstable~\cite{Dalfovo}.
We obtain
$\Omega_{\nu}/2\pi\approx 0.52\nu_{\rho}\approx 88$~Hz, which is larger than
$\Omega_c/2\pi$ but still too low to account for the experimental result.

Also shown in Fig.~\ref{omega_ani} are the frequencies of the anomalous modes
$|\omega_a|/2\pi=|\varepsilon_a|/h$ with even $z$ parity (plotted as positive
values). These are Bogoliubov excitations of the vortex state with negative
energy but positive norm, and relative angular momentum $-\hbar$. For
spherical or pancake geometries, the spectrum contains only one anomalous
mode. It describes the precession of the vortex about the 
center of the trap, in
the same sense as the circulation about the core~\cite{Linn} at a frequency
$\omega_a={\case 3/5}\Omega_c$ in the TF limit~\cite{Fetter2}. As
$\lambda$ increases and the vortex line stretches, however, additional
anomalous modes appear in the energy spectrum~\cite{Garcia2,Svidzinsky},
corresponding to the precession of a spatially deformed vortex line.
Physically, $\Omega_c$ involves only a straight vortex (note, however, that
Ref.~\onlinecite{Garcia} finds a deformed vortex for the ground state in very
elongated condensates), but $\Omega_m$ involves linearized deformations.  For
an elongated condensate ($\lambda\gg 1$), the large curvature of the
condensate surface readily induces distortions. The Bogoliubov amplitudes are
highly localized radially in the vortex core and oscillate as a function of
$z$, with maximum amplitude near the condensate surface. In the TF limit, $n$
anomalous modes appear above a critical anisotropy
$\lambda_n\geq\sqrt{n(n+1)/2}$~\cite{Svidzinsky}.

Excitation of the anomalous modes can lower the system's free energy,
destabilizing the vortex state.
Metastability of the vortex is guaranteed for rotation frequencies
exceeding the largest anomalous mode, $\Omega_m=\mbox{max}(|\omega_a|)$.
In the TF limit, $\Omega_m$ exceeds $\Omega_c$ when
$\lambda>2$~\cite{Svidzinsky}, which is close to the value of
$\lambda\approx 2.5$ found numerically
(see Fig.~\ref{omega_ani}). 
For the ENS parameters, numerical analysis
finds the
metastability critical frequency
$\Omega_m/2\pi= 0.73\nu_{\rho}\approx 124$~Hz, very close to the
observed value $\Omega_{\rm obs}/2\pi\approx 120$ Hz.

For the spherical trap relevant to the JILA experiments, the  anomalous
mode frequency in the
TF limit has the predicted form $|\omega_a|/\omega
=\case{8}{5}(\xi/R)\ln(1.96R/\xi)$,
if we neglect terms of relative order
$(\xi/R)\ln(R/\xi)\approx 0.1$, where $R=(15\eta)^{1/5}$ is the TF radius and
$\xi=1/R$ is the dimensionless healing length. We find
$|\omega_a|/2\pi=1.58\pm 0.16$~Hz, in reasonable agreement with the
observed precession frequency of $1.8\pm 0.1$~Hz. The value of $|\omega_a|$,
however, is sensitive to the number of condensate atoms ($\propto N^{-2/5}$)
and the displacement of the vortex from the trap center~\cite{Lundh}
($\propto [1-(\rho/R)^2]^{-1}$).
Interestingly, the spectrum computed numerically also contains a {\it
counter}-precessing ({\it i.e.}\ non-anomalous) mode with frequency
$3.63$~Hz and two nodes along $z$;
this may correspond to the distorted `rogue' vortices 
observed in two-component
systems~\cite{Anderson}.

To make closer contact between vortex precession
and anomalous Bogoliubov modes, the dynamics of a trapped vortex
are explored by real-time propagation of the GP equation~(\ref{gp}). 
Once the ground state has been obtained, an off-center 
vortex with counterclockwise
circulation is imposed on the condensate 
wave function at $t=0$ by the method
of phase imprinting~\cite{Denschlag}. 
The vortex is displaced by $1.57d_{\rho}$
along $\hat{x}$ from the trap origin, 
corresponding to $1.3~\mu$m and
$6.1~\mu$m for the ENS and JILA condensates, respectively. 
In both cases, the
vortices undergo counterclockwise precession, 
which is the same sense as the
circulation around the vortex core.

For the JILA parameters, the period is found to be $623\pm 3$~ms (the
uncertainty reflects the coarseness of the spatial and temporal grids),
yielding a precession frequency of $\omega_p/2\pi=1.61\pm 0.01$~Hz in excellent
agreement with the value of the predicted anomalous-mode frequency discussed
above. The phonons that are also produced by phase imprinting~\cite{Denschlag}
only weakly affect the vortex motion, and they rapidly decay into unobservable
high-frequency collective modes. The vortex itself is slightly curved, as
discussed in Ref.~\onlinecite{Svidzinsky}. No noticeable variation in the
vortex displacement from the trap center (spiraling) was found after three
full precessions.

\begin{figure}[tb]
\psfig{figure=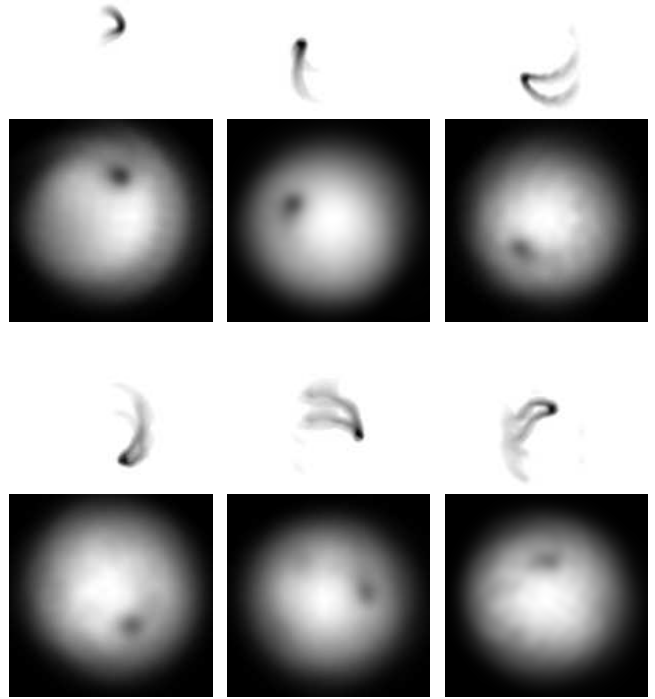,width=\columnwidth,angle=0}
\caption{Numerical results for the precession of a vortex in the ENS trap.
Frames in raster order
correspond to 4~ms increments after the initial phase imprint. Squares, which
are $8d_{\rho}\approx 6.7~\mu$m on a side, show the integrated density down
$\hat{z}$; the corresponding vortex line (viewed $9^{\circ}$ off the $z$ axis
toward $-\hat{y}$) is rendered above each frame.}
\label{precess}
\end{figure}

As shown in Fig.~\ref{precess}, 
precession of an off-center vortex in the
ENS trap is associated with a pronounced curvature of the vortex
line. The precession appears to be most rapid near the surface
of the condensate; by 12~ms, the ends of the vortex have returned to their
initial location, while the center lags behind by almost $180^{\circ}$.
Defining the precession by the motion near the surface, one obtains a
frequency of approximately 85~Hz. This value (and to some extent the shape of
the vortex) is consistent with the second-largest anomalous mode in the
excitation spectrum shown in Fig.~\ref{omega_ani},
$|\omega_a|/2\pi=0.51\nu_{\rho}\approx 87$~Hz, which has two nodes
along ${z}$. The excitation of this 
precession mode probably arises from
imprinting a circulation pattern
aligned with the $z$ axis on a condensate that is highly non-uniform
radially; the initial straight vortex would have some overlap with all of the
11 even $z$-parity negative-energy modes found above for this configuration.

\begin{figure}[tb]
\psfig{figure=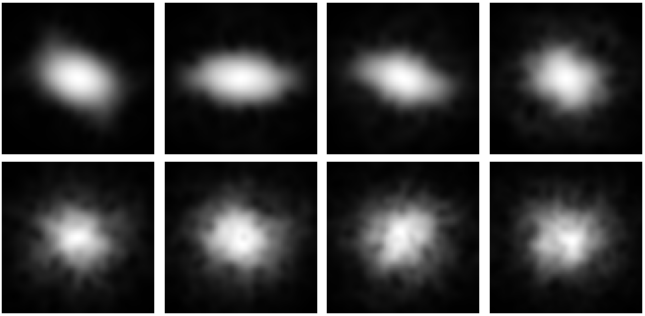,width=\columnwidth,angle=0}
\caption{The ENS condensate rotating at $\Omega=0.8\omega_{\rho}$
as a function of time. Images in raster order correspond to $150$~ms through
$500$~ms in $50$~ms increments, and are each $10\times 10~\mu$m. The
condensate density (integrated down $\hat{z}$) is shown in the rotating frame.}
\label{dynamics}
\end{figure}

The critical frequency for vortex nucleation in elongated traps is confirmed
by numerical solution of the time-dependent GP equation. The initially
axisymmetric ENS condensate is simultaneously distorted
($\epsilon_x,\epsilon_y\neq 0$) and rotated at $t=0$. The GP equation is
propagated in the rotating frame for $500$~ms with either
$\Omega=0.7\omega_{\rho}$ or $0.8\omega_{\rho}$. No vortices are produced for
the $\Omega=0.7\omega_{\rho}$ case. As shown in Fig.~\ref{dynamics}, however,
when $\Omega=0.8\omega_{\rho}$ vortices appear at the condensate surface by
$t=200$~ms, and fully penetrate the cloud by $t=300$~ms. In the absence of
dissipation, the vortex motion remains extremely turbulent; the dynamics of
nucleating vortices is discussed in greater detail elsewhere~\cite{Feder3}.
The ENS condensates are held in the rotating trap for $500$~ms, presumably
allowing the vortex structures to approach equilibrium. The simulations
indicate that dissipative processes not included in the present study, such as
the interaction of the condensate with the thermal cloud~\cite{Fedichev}, are
important in the formation of the vortex arrays observed experimentally.

For an anisotropic trap with $\epsilon_x\neq\epsilon_y$, the anomalous mode
$\omega_a(\Omega)$ becomes imaginary in the range
$|\omega_a|-\delta\lesssim\Omega\lesssim|\omega_a|+\delta$, where $2\delta =
|\epsilon_x-\epsilon_y|$~\cite{Svidzinsky}. The onset of
metastability  in such traps occurs only when $\Omega$  exceeds $|\omega_a|$
by the appropriate amount, which may help explain the relevance of
$\Omega_m\gtrsim|\omega_a|$ for the ENS experiments.

In summary, anomalous modes are interpreted as defining both the vortex
precession in the JILA experiments~\cite{Anderson} and the critical frequency
for the appearance of the first vortex in the ENS experiments~\cite{Dalibard}.

\begin{acknowledgments}
The authors are grateful to B.~P.~Anderson, E.~A.~Cornell, J.~Dalibard,
J.~Denschlag, and S.~L.~Rolston for numerous fruitful discussions. This work
was supported in part by the U.S.\ office of Naval Research, by NSF Grant
No.~DMR 99-71518, and by Stanford University.
\end{acknowledgments}

\end{document}